\documentclass[%
 aip,
 amsmath,amssymb,
preprint,%
 reprint,%
]{revtex4-1}

\usepackage[text=PREPRINT]{draftwatermark}
\usepackage{graphicx}
\usepackage{dcolumn}
\usepackage{bm}

\usepackage[utf8]{inputenc}
\usepackage[T1]{fontenc}
\usepackage{mathptmx}
\usepackage{etoolbox}

\makeatletter
\def\@email#1#2{%
 \endgroup
 \patchcmd{\titleblock@produce}
  {\frontmatter@RRAPformat}
  {\frontmatter@RRAPformat{\produce@RRAP{*#1\href{mailto:#2}{#2}}}\frontmatter@RRAPformat}
  {}{}
}%
\makeatother
\begin{document}

\preprint{AIP/123-QED}

\title{Universal scaling of shear thickening suspensions under acoustic perturbation}
\author{Anna R. Barth}
\author{Navneet Singh}
\author{Stephen J. Thornton}
\author{Pranav Kakhandiki}
\author{Edward Y.X. Ong}
\affiliation{ 
Department of Physics, Cornell University, Ithaca, New York 14850, USA
}
\author{Meera Ramaswamy}
\affiliation{Department of Mechanical Engineering, University of Minnesota, Minneapolis, Minnesota 55455, USA}
\author{Abhishek M. Shetty}
 \affiliation{ 
Department of Rheology, Anton Paar, Ashland, Virginia 23005, USA
}
\author{Bulbul Chakraborty}
\affiliation{Department of Physics, Brandeis University, Waltham, Massachusetts 02453, USA}
\author{James P. Sethna}
\author{Itai Cohen}
\affiliation{ 
Department of Physics, Cornell University, Ithaca, New York 14850, USA
}

\date{\today}

\begin{abstract}
Tuning shear thickening behavior is a longstanding problem in the field of dense suspensions. Acoustic perturbations offer a convenient way to control shear thickening in real time, opening the door to a new class of smart materials. However, complete control over shear thickening requires a quantitative description for how suspension viscosity varies under acoustic perturbation. Here, we achieve this goal by experimentally probing suspensions with acoustic perturbations and incorporating their effect on the suspension viscosity into a universal scaling framework where the viscosity is described by a scaling function, which captures a crossover from the frictionless jamming critical point to a frictional shear jamming critical point. Our analysis reveals that the effect of acoustic perturbations may be explained by the introduction of an effective interparticle repulsion whose magnitude is roughly equal to the acoustic energy density. Furthermore, we demonstrate how this scaling framework may be leveraged to produce explicit predictions for the viscosity of a dense suspension under acoustic perturbation. Our results demonstrate the utility of the scaling framework for experimentally manipulating shear thickening systems.
\end{abstract}

\maketitle

\section{Introduction}
Many dense colloidal and granular suspensions shear thicken, exhibiting a viscosity which increases with shear stress, sometimes quite dramatically \cite{brown2010shearthickening, brown2014shearthickening, wagner2009shearthickening}. Such suspensions have found a variety of applications from industrial polishing to protective fabrics to impact-resistant batteries \cite{zarei2020applications, gurgen2017applications, wei2022applications}. However, shear thickening is also responsible for a litany of engineering challenges. Shear thickening increases the power required to mix and transport dense suspensions, and in the most severe cases, can even lead to equipment damage \cite{brown2014shearthickening,barnes1989shearthickening}. Therefore there is an interest in reducing, tuning, and controlling shear thickening behavior. Many mechanisms exist to reduce shear thickening by modifying the materials that compose the suspension, for example by reducing the volume fraction of solid particles or modifying surface properties of the particles \cite{yahia2014cement,Toussaint2009cement,Bossis2017surfacechemistry,mckinley2022adhesion, james2018adhesion_Hbonding, james2019adhesion_Hbonding}. However, for many practical applications, the material cannot be modified. In such circumstances, it is desirable to tune the viscosity with an external control knob. 

One way to control shear thickening behavior is by applying acoustic perturbations. Acoustic perturbations have been shown to dethicken and even unjam shear thickening suspensions \cite{sehgalAcoustics,edwardMemory}. Upon application of the acoustic field, the suspension transitions to a lower-viscosity state, and the viscosity recovers once the field is removed.  This rapid, reversible response allows for tuning of the suspension's viscosity in real time. Additionally, unlike other techniques for tuning suspension rheology in real time, such as magnetorheology \cite{morillas2020magnetorheo} and orthogonal shear perturbations \cite{neil2016OSP,meeraOSP}, acoustic perturbations can be easily implemented in various flow geometries simply by installing a piezoelectric transducer. These advantages make acoustic perturbations a promising tool for a wide variety of applications involving dense suspensions.

Past work \cite{sehgalAcoustics,sehgal2024metamaterials} has shown that stronger acoustic perturbations result in a greater dethickening effect. There has yet to be, however, a systematic study of how the magnitude of this effect depends on other determinants of suspension viscosity. Here, we work towards that goal by varying the strength of acoustic perturbation, suspension volume fraction, and applied shear stress to map out how these factors collectively determine the suspension viscosity. 
We analyze these measurements by building on a recently developed universal scaling framework, where shear thickening is controlled by two jamming critical points: one frictionless and one frictional \cite{meeraScaling,meeraOSP}. From this perspective, the viscosity of a shear thickening suspension can be described by a crossover scaling function whose argument is a scaling variable that takes into account the combined contributions of shear stress and volume fraction. Here, we investigate whether the effects of acoustic perturbations can folded into this universal scaling framework. Because this framework is grounded in critical behavior, the crossover scaling function is universal across shear thickening suspensions composed of different materials. Thus, this approach could offer a complete quantitative description of suspension rheology under acoustic perturbations so that with only a few measurements in any particular system, one could map out the entire parameter space. 

\section{Methods}

Suspensions were prepared from polydisperse aluminosilicate ceramic microspheres with a size range of $1-15\,\mu m$ (Nationwide Protective Coating, Inc.) mixed with glycerol (Ricca Chemical). Quantities of dry particles and glycerol were measured by mass in order to estimate the volume fraction of each sample, taking the density of glycerol to be $\rho_s=1.26\,$g/mL and the density of particles to be $\rho_p=2.4\,$g/mL. Despite the density mismatch between solvent and particles, sedimentation was not a significant issue under the conditions used for measurements \cite{supp_stability}. Suspensions were mixed by hand, vortexed, and sonicated for 30 minutes immediately before beginning each experiment.

\begin{figure}
    \centering
    \includegraphics[width=\linewidth]{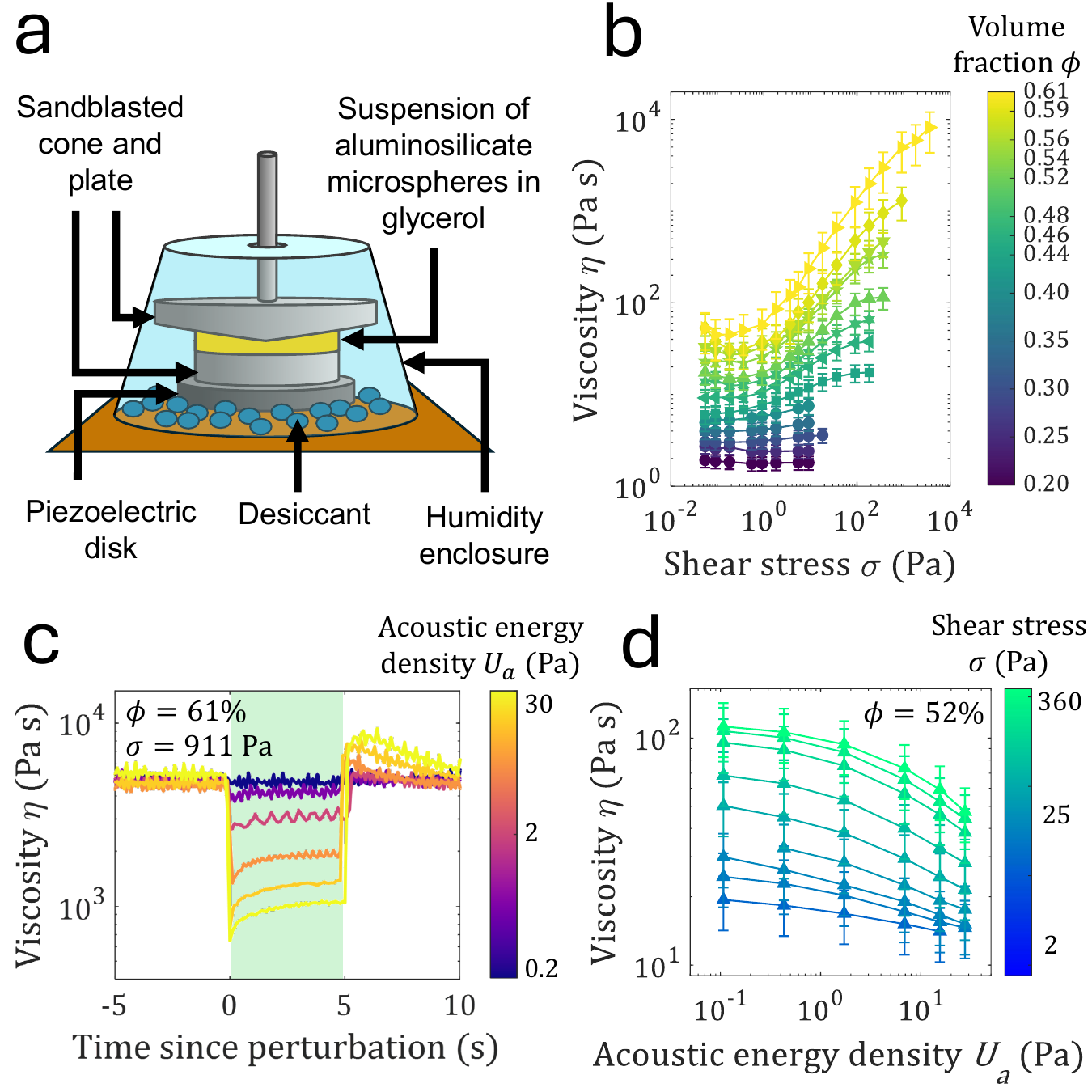}
    \caption{Suspensions dethicken under acoustic perturbations. (a) Schematic of apparatus used to apply acoustic perturbations on the rheometer. (b) Viscosity $\eta$ versus shear stress $\sigma$ for various volume fractions $\phi$ of aluminosilicate microspheres in glycerol. (c) Viscosity versus time for a few example acoustic perturbations, applied on a suspension with volume fraction $\phi=0.61$ and under applied shear stress $\sigma=911\,$Pa. The color shows the acoustic energy density $U_a$ of each perturbation. (d) Viscosity versus acoustic energy density $U_a$ for a volume fraction of $\phi=0.52$. The applied shear stress $\sigma$ is shown by the color. }
    \label{fig:rawdata}
\end{figure}

Measurements were performed on an Anton Paar MCR 702 rheometer with a cone and plate geometry. In order to prevent the glycerol from absorbing water from the air, the sample was surrounded by calcium sulfate desiccant (Drierite) and kept under a plastic enclosure (Fig~\ref{fig:rawdata}a). Under these conditions, the viscosity of the suspension was stable in time, varying by less than 15\% over the time scale of the experiment \cite{supp_stability}. Stress sweeps for each sample demonstrate clear shear thickening behavior (Fig.~\ref{fig:rawdata}b).

Acoustic perturbations were applied using a custom bottom plate geometry with a piezoelectric disk (APC International) driven with an AC voltage at its resonant frequency $f=1.15\,$MHz to supply acoustic perturbations, as previously described in \cite{sehgalAcoustics,edwardMemory}. The amplitude of the perturbation was controlled by the magnitude of the AC voltage driving the piezo. We quantify the strength of these perturbations by the acoustic energy density $U_a$ \cite{garat2022vibrations,hanotin2015}, calculated as
\begin{equation}
    U_a=\frac12\rho_0 \delta_a^2(2\pi f)^2
\end{equation}
where $\rho_0$ is the density of the medium, and $\delta_a$ and $f$ are the amplitude and frequency of the acoustic oscillations. Here, we take the density of the medium to be the average density of the solvent and the solid particles $\rho_0=\frac12(\rho_s+\rho_p).$ We estimate the amplitude of the acoustic oscillations $\delta_a=Vd_{33}$, where $V$ is the amplitude of the AC voltage attached to the piezo and $d_{33}$ is the piezoelectric modulus. For our experiments, $V$ ranged from $0\,$V to $80\,$V and $d_{33}=3\times10^{-10}\,$m/V, as reported by the manufacturer (APC International, material 841).


During experiments, a constant shear stress $\sigma$ was applied via the rheometer, and then acoustic perturbations were applied for 5 seconds each.  A few representative examples of responses to acoustic perturbations are shown in Fig.~\ref{fig:rawdata}c. At the onset of the acoustic perturbation, the suspension viscosity sharply decreased. When the perturbation was removed, the viscosity showed a rapid increase and a more gradual return to its original value. The system was monitored manually to ensure the waiting time between measurements was sufficient for this recovery. 
To quantify the effect of these acoustic perturbations, we report the average viscosity for the period where the perturbations were applied. The viscosity's dependence on acoustic energy density and applied shear stress for one volume fraction ($\phi=0.52$) is shown in Fig.~\ref{fig:rawdata}d.




\section{Results}


\label{sec:scaling_collapse}

We begin by examining the viscosity $\eta$ of the suspensions in the absence of an acoustic field (i.e., the data shown in Fig.~\ref{fig:rawdata}b). Past work \cite{meeraScaling,meeraOSP} recognized that the viscosity of a shear thickening suspension can be related to its volume fraction $\phi$ and applied shear stress $\sigma$ through a Widom-like scaling function \cite{widom1965}, inspired by the Wyart-Cates model \cite{wyartcates}:

\begin{equation}
     \eta(\phi,\sigma) =(\phi_0-\phi)^{-2}\mathcal{F}\left(\frac{C(\phi)e^{-\sigma^*_0/\sigma}}{\phi_0-\phi}\right).
     \label{eqn:scaling_ansatz}
\end{equation}
Here, $\mathcal{F}$ is a crossover scaling function that captures the transition from frictionless jamming to frictional jamming. $\phi_0$ is the frictionless jamming volume fraction, a material-dependent parameter related to the shape and size distribution of the particles. As in the Wyart-Cates model, $\sigma^*_0$ is the shear stress required to overcome interparticle repulsion and form frictional contacts, and $f(\sigma)=e^{-\sigma^*_0/\sigma}$ represents the fraction of interparticle contacts that are frictional rather than lubricated \cite{wyartcates}. $C(\phi)$ is the anisotropy factor, a function of volume fraction hypothesized to quantify the fraction of interparticle contacts which lie along the compressive axis and contribute to shear thickening \cite{meeraScaling}. Equation~\ref{eqn:scaling_ansatz} identifies two renormalization group flow directions originating at the frictionless jamming point. One of these directions, $\phi_0 - \phi$, simply corresponds to the distance in volume fraction to the frictionless jamming point. The other, $C(\phi)e^{-\sigma^*_0/\sigma}$, corresponds to the contribution of friction to the viscosity.\footnote{Intriguingly, the scaling variable $x=e^{-\sigma^*_0/\sigma}/(\phi_0-\phi)$ is precisely what would arise from stress $\sigma$ being a marginally relevant variable in the renormalization group flow about the frictionless jamming point at $(\phi=\phi_0,\sigma=0)$ (a derivation of this result can be found in the Supplementary Material). How, then, can we interpret $C(\phi)$ as it appears in Equation~{\ref{eqn:scaling_ansatz}}? $C(\phi)$ may be interpreted as an analytic correction to scaling that arises due to higher order terms in the renormalization group flow of $\sigma.$ Indeed, our scaling variable may be rewritten in a form where analytic corrections are incorporated directly into a transformation of the stress, $x=e^{-\sigma^*_0/\Sigma(\phi,\sigma)}/(\phi_0-\phi)$. Here, the transformed stress is $\Sigma(\phi,\sigma)=\sigma\times\sigma^*_0/(\sigma^*_0-\sigma\ln C(\phi))$.}

\begin{figure}
    \centering
    \includegraphics[width=\linewidth]{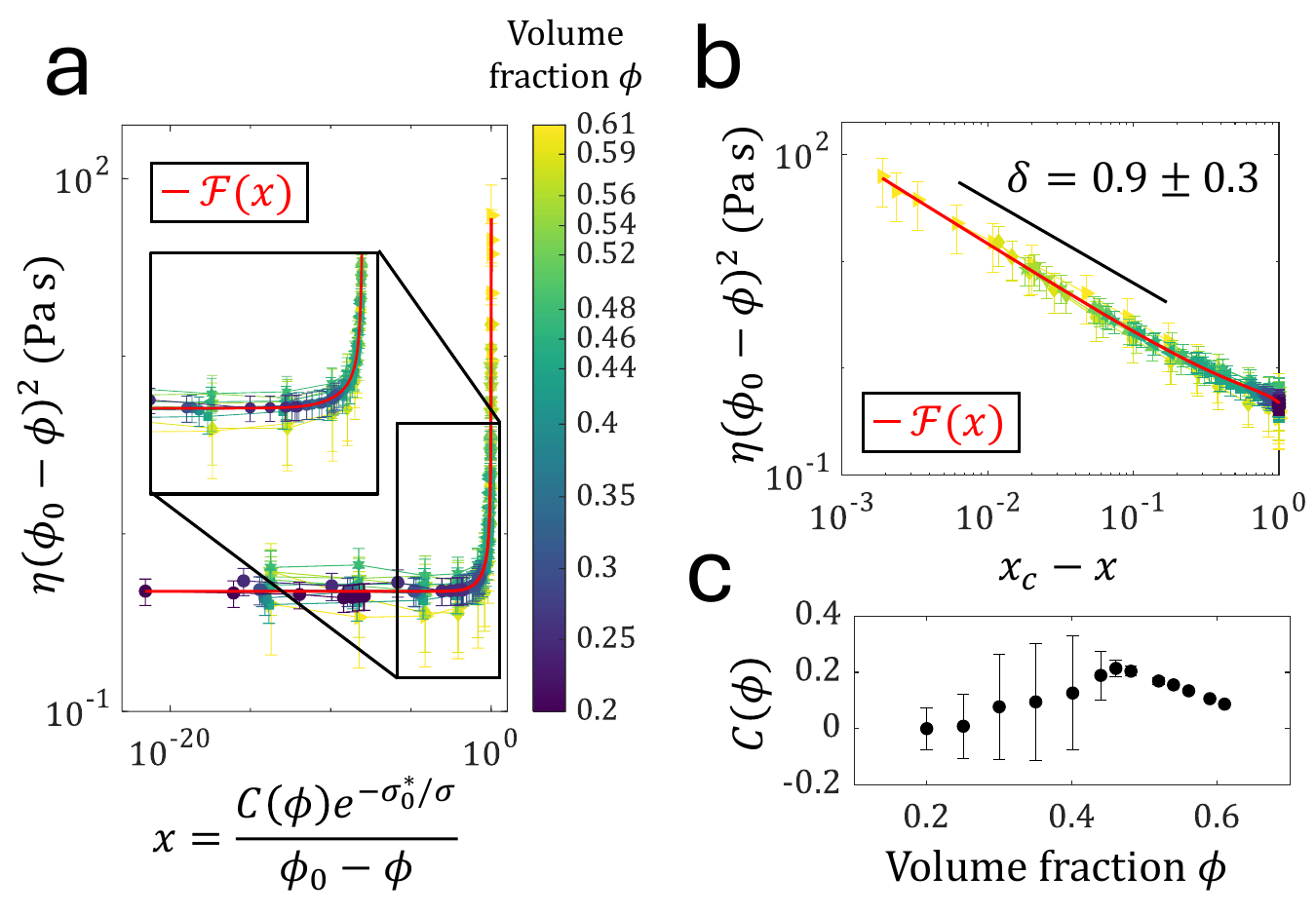}
    \caption{Shear thickening is a precursor to shear jamming and is controlled by a crossover scaling function. (a) Viscosity data in the absence of an acoustic field ($U_a=0$) plotted as $\eta(\phi_0-\phi)^2$ versus $x= \frac{C(\phi)e^{-\sigma^*_0/\sigma}}{\phi_0-\phi}.$ As predicted by Equation~\ref{eqn:scaling_ansatz}, the data collapse onto a single curve, $\mathcal{F}(x)$, whose shape is approximated by a fitting function shown as a red line. The inset shows the part of the scaling function near $x=x_c=1$, where the scaling function diverges.  (b)  $\eta(\phi_0-\phi)^2$ versus $x_c-x$, highlighting that $\mathcal{F}(x)$ diverges at $x_c$ as $\mathcal{F}\sim(x_c-x)^{-\delta}$ where $\delta=0.9\pm0.3$. (c) $C(\phi)$, the anisotropy factor plotted against volume fraction $\phi$.}
    \label{fig:acous_free}
\end{figure}

Equation~\ref{eqn:scaling_ansatz} implies that when $\eta(\phi_0-\phi)^2$ is plotted against $x = C(\phi)e^{-\sigma^*_0/\sigma}/(\phi_0-\phi),$ all data across different volume fractions and shear stresses should collapse onto a single curve $\mathcal{F}(x)$ \cite{meeraScaling,meeraOSP}. We seek such a collapse of the acoustics-free data by varying the parameters involved in $x$ ($C(\phi)$, $\sigma^*_0$, and $\phi_0$) according to a fitting procedure described in the Supplementary Material \cite{supp_fitting}. The resulting data collapse along with the best fit $\mathcal{F}(x)$ curve are shown in Fig.~\ref{fig:acous_free}a and Fig.~\ref{fig:acous_free}b. 

Analysis of the scaling function $\mathcal{F}(x)$ shows it is flat for low values of $x,$ corresponding to the low-stress regime where the viscosity controlled by the frictionless jamming point \cite{meeraScaling,wyartcates}, following $\eta\sim(\phi_0-\phi)^{-2}.$ Near $x_c$, corresponding to the high-stress regime where the viscosity is controlled by the frictional jamming point, $\mathcal{F}(x)$ diverges as a power law, $\mathcal{F}(x)\sim(x_c-x)^{-\delta}$ as shown in Fig.~\ref{fig:acous_free}b. Note that because $x$ contains a multiplicative factor $C(\phi)$, there is an overall scale that may be chosen for $x$; we choose this scale so that $x_c=1$. We find the value of the exponent $\delta$ by fitting a functional form to $\mathcal{F}(x)$ \cite{supp_fitting} and find $\delta=0.9\pm0.3.$ Thus, as $x$ approaches a critical value $x_c=1$, the viscosity is controlled by the frictional jamming point\cite{meeraScaling, supp_eta_scaling} and follows $\eta\sim(\phi_J-\phi)^{-0.9}$.

The anisotropy factor $C(\phi)$ is shown in Fig~\ref{fig:acous_free}c. $C(\phi)$ is non-monotonic in volume fraction, consistent with past work on other suspensions \cite{meeraScaling,meeraOSP}. We note that, for low volume fractions, the confidence interval on $C(\phi)$ is rather large. This reflects the fact that low volume fractions correspond to small values of $x$. Since $\mathcal{F}(x)$ is nearly flat in this region, relatively large changes in $x$ correspond to very little change in $\mathcal{F}(x)$. Therefore for lower volume fractions, there is a wide range of $C(\phi)$ over which the quality of the data collapse is still quite good.

\begin{figure}
    \centering
    \includegraphics[width=\linewidth]{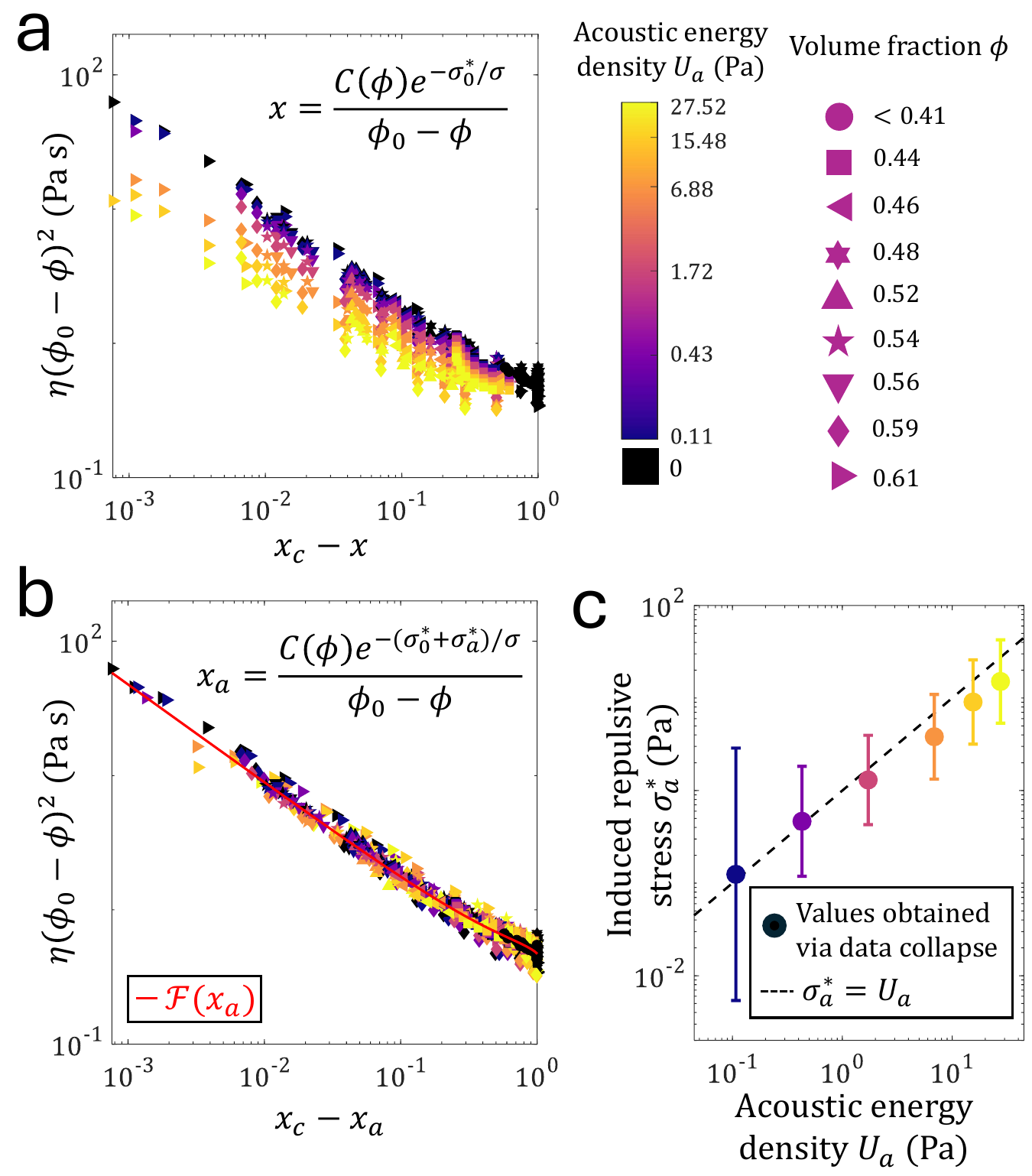}
    \caption{The effect of acoustic perturbations may be incorporated into a universal scaling framework. (a) $\eta(\phi_0-\phi)^2$ plotted against $x_c-x$ before incorporating the effects of acoustic perturbations. Acoustic energy density $U_a$ is shown by the color bar. Volume fraction is shown by symbols (same symbols as Fig.~\ref{fig:rawdata}b). The viscosity data do not collapse onto a single curve. Error bars are excluded for visual clarity. (b) After incorporating the effects of acoustic perturbations into the scaling variable $x_a$, the data now collapse onto the curve $\mathcal{F}(x_a)$. (c) The acoustic contribution $\sigma^*_a$ to the interparticle repulsive stress versus acoustic energy density $U_a$. The circles are the values that produce the best collapse of the viscosity data, and the dotted line is $\sigma^*_a=U_a.$}
    \label{fig:acous_collapse}
\end{figure}

Having analyzed the data with no acoustic perturbations (i.e. acoustic energy density $U_a=0$), we now seek a strategy to incorporate the effect of acoustic perturbations into this framework. When data with acoustic perturbations (i.e. nonzero $U_a$) are plotted as $\eta(\phi_0-\phi)^2$ versus $x$, because of the acoustic dethickening effect, the data no longer collapse, as shown in Fig.~\ref{fig:acous_collapse}a. We hypothesize that the acoustic perturbations induce some additional interparticle repulsive stress $\sigma^*_a$, which could be supplied, for example, by microstreaming effects \cite{fabre2017microstreaming,melody2023microstreaming} or by a thermal-like interaction arising from diffusive particle motion \mbox{\cite{garat2022vibrations,gaudel2017vibrations}}. We also conjecture that the acoustic repulsion adds onto the pre-existing interparticle repulsion, so that the total repulsive barrier $\sigma^*_\text{total}$ is:
\begin{equation}
    \sigma^*_\text{total}=\sigma^*_0+\sigma^*_a(U_a).
\end{equation}
Accordingly, Equation~\ref{eqn:scaling_ansatz} becomes:
\begin{equation}
     \eta(\phi,\sigma,U_a) =(\phi_0-\phi)^{-2}\mathcal{F}\left(\frac{C(\phi)e^{-(\sigma^*_0+\sigma^*_a(U_a))/\sigma}}{\phi_0-\phi}\right).
     \label{eqn:scaling_ansatz_acous}
\end{equation}

As before, we use a least-squares fitting procedure to determine the values of the parameters $\phi_0$, $C(\phi)$, $\sigma^*_0$, and $\sigma^*_a(U_a)$ \cite{supp_fitting}. This approach produces an excellent collapse of the data across all volume fractions, shear stresses, and strengths of acoustic perturbations, as shown in Fig.~\ref{fig:acous_collapse}b. The successful collapse is consistent with the hypothesis that acoustic perturbations provide an additional effective repulsion between particles that must be overcome by the shear stress to facilitate frictional interactions that generate thickening. 


Finally, we find that the acoustic contribution to the interparticle repulsive stress $\sigma^*_a(U_a)$ is approximately equal to the acoustic energy density, that is, $\sigma^*_a\sim U_a$ (see Fig.~\ref{fig:acous_collapse}c). This relation, which can be argued from dimensional analysis, is to be expected for any interparticle interaction driven by acoustics. Collectively, these results indicate that acoustic perturbations can be easily folded into the universal framework organizing thickening transitions. 

\begin{figure}
    \centering
    \includegraphics[width=\linewidth]{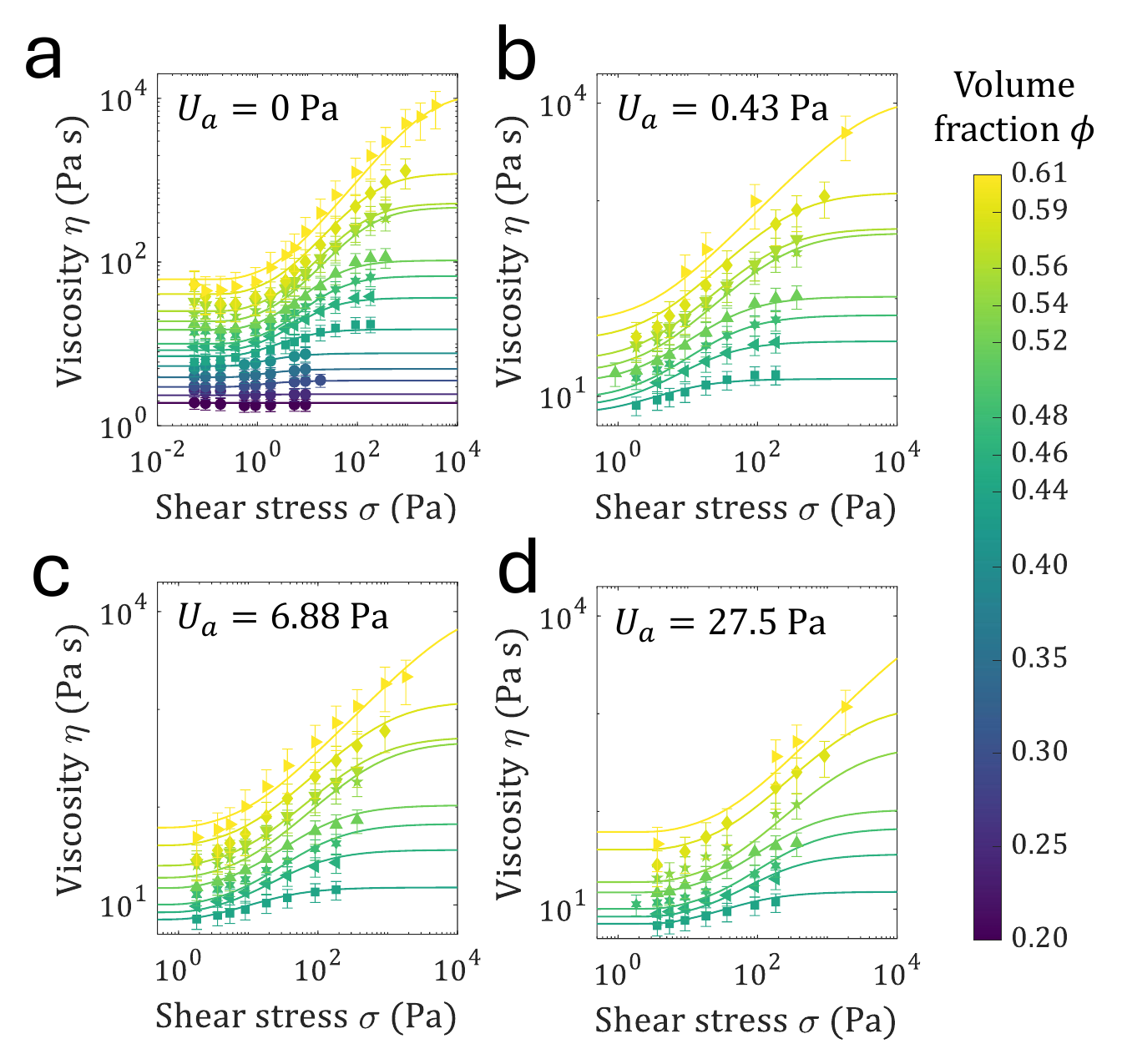}
    \caption{Comparison between experimental measurements (symbols) and predictions of the scaling framework (lines) for suspension viscosity $\eta$ as a function of volume fraction $\phi$, shear stress $\sigma$, and acoustic energy density $U_a,$ for $U_a=0$, $0.43$ Pa, $6.88$ Pa, and $27.5$ Pa. Results for other acoustic energy densities can be found in the Supplementary Material.}
    \label{fig:viscosity_predictions_main}
\end{figure}

This scaling framework provides a complete theory for how changes in the viscosity arise from the combined contributions of volume fraction, shear stress, and acoustic perturbations. As such it can be used practically to predict the effect of acoustic perturbations on the viscosity of a shear thickening suspension. We explicitly generate these predictions from Equation~{\ref{eqn:scaling_ansatz_acous}}. This calculation requires a functional form for the scaling function $\mathcal{F}(x)$. We infer this functional form by fitting to the shape of the collapsed data. The resultant function is shown as a red line in Fig.~{\ref{fig:acous_collapse}b} \cite{supp_fitting}. Using this functional form, along with the parameters discovered in the data collapse procedure, we observe excellent agreement between the predictions of Equation~{\ref{eqn:scaling_ansatz_acous}} and the experimental dataset, as can be seen in Fig.~{\ref{fig:viscosity_predictions_main}}.


\section{Discussion}

We find that the effect of acoustic perturbations can be incorporated into a universal scaling framework for shear thickening by simply modifying $\sigma^*$ to account for an additional effective interparticle repulsive barrier. Therefore we propose that suspensions dethicken under acoustic perturbation because the acoustic field induces a stronger interparticle repulsion, causing some particles to switch from frictional contact to a lubricated interaction, thereby facilitating flow and lowering the viscosity.
Furthermore, we find that the additional repulsive stress $\sigma^*_a$ induced by the acoustic perturbation is nearly equal to the average acoustic energy density $U_a$. 

It is interesting to speculate about possible underlying mechanisms that provide this effective interparticle repulsion for acoustic perturbations. For example, microstreaming, the creation of vortical flows near the particle surface, is known to induce interparticle repulsion \cite{melody2023microstreaming,fabre2017microstreaming}. This repulsion is controlled by a Stokes number $\Omega=(2\pi f)R^2\rho_s/\eta_s$ where $f$ is the acoustic frequency, $R$ is the particle radius, and $\rho_s$ and $\eta_s$ are the density and viscosity of the solvent. For our system, which includes a range of particle sizes, $\Omega$ ranges from roughly 0.01 to 2. These values are small enough that microstreaming forces would be expected to supply substantial repulsion \cite{fabre2017microstreaming}. Investigating this mechanism further would be an interesting avenue for future research.

Our results also draw a connection between these acoustic perturbations and previously-studied vibrations. Past work found  that lower-frequency (10-50 Hz) vibrations can unjam dense suspensions by supplying an effective interparticle repulsive stress roughly equal to the vibration energy density  \cite{garat2022vibrations}. Thus, across a wide range of frequencies, the unjamming and dethickening effect of mechanical vibrations can be captured by this effective repulsion model. It is interesting to contemplate how orthogonal shear perturbations \cite{meeraOSP,neil2016OSP} might fit into this picture. Orthogonal shear perturbations are also a mechanical oscillation, but they are typically considered a boundary effect, while vibrations are a bulk effect, so these two types of perturbation have been assumed to operate via different mechanisms. It may be possible that orthogonal shear perturbations can also be explained by an effective repulsion, unifying a wide range of mechanical perturbations under a single conceptual umbrella. Future work might explore whether such effective force models can also explain other means of tuning the viscosity of a shear thickening suspension, such as adhesion \cite{mckinley2022adhesion, james2018adhesion_Hbonding, james2019adhesion_Hbonding} or the introduction of activity, for example, via Quincke rotation \cite{lobry1999quincke, pannacci2007quincke, lemaire2008quincke, edwardThesis}.

The strategy we employ here to describe suspension rheology is grounded in the language of critical phenomena, which has become an increasingly powerful tool for investigating soft materials, ranging from lipid membranes \cite{keller2008membranes,veatch2007lipids} to cartilage \cite{thomas2022cartilage,fred2016cartilage} to 
biomolecular condensates \cite{eric2024droplets,brangwynne2009pgranule}. Therefore our approach for describing the effect of acoustics may be useful for other soft materials with tunable mechanical properties. Recent work has shown that acoustic perturbations may also be used to tune the elastic properties of colloidal gels \cite{gibaud2020rheoacoustic_gels}. We suggest that a similar universal scaling framework might be a useful tool for modeling the properties of these gels under acoustic perturbation, particularly since it is already known that the rheology of soft amorphous solids exhibits critical behavior \cite{suman2020gel_universality, ohern2003jamming, silbert2005jamming, ellenbroek2006jamming, lerner2012unifying_gels_and_suspensions}. 

Finally, we have shown that this scaling framework has practical value as a model for the effect of acoustic perturbations; we have demonstrated this explicitly by generating predictions for the suspension viscosity at different acoustic energy densities. Because this framework is based on a universal scaling function $\mathcal{F}$, this framework could be used for any shear thickening suspension under acoustic perturbation. One would simply require a few initial measurements of $\eta(\phi,\sigma,U_a)$ to characterize particular system-specific details: the analysis in Section~\ref{sec:scaling_collapse} must be followed to obtain the non-universal quantities $\phi_0$, $C(\phi)$, $\sigma^*_0$, and $\sigma^*_a(U_a)$. Then, Equation~\ref{eqn:scaling_ansatz_acous} provides a complete description of the rheology with or without acoustics. Thus, our analysis serves as a blueprint for characterizing the effect of acoustic perturbations on any dense suspension, resulting in a quantitative model which can be used to calculate the acoustic energy density required to accomplish a given target viscosity. This capability is an important step on the path to fluid metamaterials with tunable viscosities. 




\section*{Acknowledgments}
We thank Christopher Ness, Jason Z. Kim, Melody X. Lim, and Shreyas Sudhaman for valuable discussions. We also acknowledge Anton Paar for use of the MCR 702 rheometer through their VIP academic research program. This material is based upon work supported by the National Science Foundation Graduate Research Fellowship under Grant No. DGE-2139899. A.R.B., N.S., S.J.T., P.K., E.Y.X.O., M.R., I.C., and J.P.S. were supported by NSF DMR-2327094.

\section*{Author declarations}
The authors have no conflicts of interest to disclose.

\section*{Data availability statement}
The data that support the findings in this study are available from the corresponding author upon reasonable request.

\nocite{*}
\bibliography{apssamp}

\end{document}